\documentstyle[preprint,aps]{revtex}
\tightenlines

\begin{document}

\title{Cosmic acceleration as the solution to the cosmological 
constant problem\footnote{astro-ph/9910093 v3, May 1, 2001}}

\author{\normalsize{Philip D. Mannheim} \\
\normalsize{Department of Physics,
University of Connecticut, Storrs, CT 06269} \\
\normalsize{mannheim@uconnvm.uconn.edu} \\}

\maketitle

\begin{abstract}

In this paper we provide both a diagnosis and resolution of the cosmological 
constant problem, one in which a large (as opposed to a small) cosmological 
constant $\Lambda$ can be made compatible with observation. We trace the 
origin of the cosmological constant problem to the assumption that the local
gravitational Newton constant $G$ (as measured in a Cavendish experiment) 
sets the scale for global cosmology. And then we show that once this
assumption is relaxed, the very same cosmic acceleration which has served to
make the cosmological constant problem so very severe can instead then
serve to provide  us with its potential resolution. In particular, we
present an alternate cosmology, one based on conformal gravity, a theory
whose effective cosmological $G_{eff}$ not only differs from the Cavendish
one by being altogether smaller than it, but, by even being explicitly
negative, naturally leads to cosmological repulsion. We show in the
conformal theory, that once given only that the sign of $\Lambda$ is
specifically the negative one associated with spontaneous scale symmetry
breaking, then, that alone, no matter how big $\Lambda$ might actually be in
magnitude, is sufficient to not only make the actually measurable
contribution $8\pi G_{eff}\Lambda/3cH^2(t_0)$ of $\Lambda$ to current era
cosmology naturally be of order one today, but to even do so in a way which
is fully compatible with the recent high $z$  supernovae cosmology data.
Thus to solve the cosmological constant problem we do not need to change or
quench the energy content of the universe, but rather we only need change
its effect on cosmic evolution.

\end{abstract}

\section{Diagnosis of the Cosmological Constant Problem}

The recent discovery \cite{Riess1998,Perlmutter1998} of a cosmic acceleration
has made the already extremely disturbing cosmological constant problem even 
more vexing than before. Specifically, a phenomenological fitting to the new 
high $z$ supernovae Hubble plot data using the standard Einstein-Friedmann 
cosmological evolution equations	
\begin{equation}
\dot{R}^2(t) +kc^2=\dot{R}^2(t)(\Omega_{M}(t)+\Omega_{\Lambda}(t))
\label{1}
\end{equation}
\begin{equation}
\Omega_{M}(t)+\Omega_{\Lambda}(t)+\Omega_{k}(t)=1
\label{1l}
\end{equation}
\begin{equation}
q(t)=(n/2-1)\Omega_{M}(t)-\Omega_{\Lambda}(t)
\label{1h}
\end{equation}
where $\Omega_{M}(t)=8\pi G\rho_{M}(t)/3c^2H^2(t)$ is due to ordinary matter 
(viz. matter for which $\rho_{M}(t)=A/R^n(t)$ where $A>0$ and $3\leq n \leq 
4$), where $\Omega_{\Lambda}(t)=8\pi G\Lambda/3cH^2(t)$ is due to a
cosmological constant $\Lambda$ and where $\Omega_{k}(t)=-kc^2/\dot{R}^2(t)$
is due to the spatial 3-curvature $k$, has revealed that not only must the
current era $\Omega_{\Lambda}(t_0)$ actually be non-zero today, it is even
explicitly  required to be of order one. Typically, the allowed parameter
space compatible with the available data is found to be centered on the line 
$\Omega_{\Lambda}(t_0)=\Omega_{M}(t_0)+1/2$ or so with (the presumed 
positive) $\Omega_{M}(t_0)$ being found to be limited to the range $(0,1)$
and $\Omega_{\Lambda}(t_0)$ to the range $(1/2,3/2)$ or so, with the current
($n=3$) era deceleration parameter $q(t_0)=(n/2-1)\Omega_{M}(t_0) -
\Omega_{\Lambda}(t_0)$ thus having to approximately lie within the
$(-1/2,-1)$interval.\footnote{The spread around the line 
$\Omega_{\Lambda}(t_0)=\Omega_{M}(t_0)+1/2$ is of order $\pm 1/2$, to thus
allow solutions in which the current era $\Omega_{M}(t_0)$  is negligible,
with $\Omega_{\Lambda}(t_0)$ having to then lie in the $(0,1)$ interval and
$q(t_0)$ in $(0,-1)$. (For explicit acceptable $\Omega_{M}(t_0)=0$ fits in 
the 'empty universe' case where $\Omega_{\Lambda}(t_0)$ is also zero see
\cite{Dev2000}). While completely foreign to the standard model, as we shall
see below, universes where $\rho_{M}(t_0)$ has little effect on current era
cosmic evolution can nonetheless actually occur quite naturally in the
alternate conformal gravity theory which we explore in this paper.} Thus,
not only do we find that the universe is currently accelerating, but
additionally we see that with there being no allowed
$\Omega_{\Lambda}(t_0)=0$ solution at all (unless $\Omega_{M}(t_0)$ could
somehow be allowed to go negative), the longstanding problem (see e.g.
\cite{Weinberg1989,Ng1992} for some recent reviews) of trying to find some
way by which $\Omega_{\Lambda}(t_0)$ (and thus $q(t_0)$) could be quenched
by many orders of magnitude from its quantum gravity Planck temperature
expectation or its typical $c|\Lambda|=\sigma T_V^4$ ($T_V\simeq
10^{16}~^{\circ}K$) particle physics expectation has now been  replaced by
the need to find a specific such mechanism which in practice (rather than
just in principle) would explicitly put $\Omega_{\Lambda}(t_0)$ into this
very narrow
$(1/2,3/2)$ box. Not only is it not currently known how it might be possible
to actually do this, up to now no mechanism has been identified which might
even fix the sign of the standard model $\Omega_{\Lambda}(t_0)$ let alone
its magnitude. 

Now while such quenching of $\Lambda$ has yet to be achieved, it is 
important to note that even if it were to actually be achieved, the very use
of such a quenched $\Lambda$ in Eq. (\ref{1}) then creates a further problem
for the standard model, one related to the initial conditions of the then
associated early universe. Specifically, with both the current era 
$\Omega_{M}(t_0)$ and a such quenched $\Omega_{\Lambda}(t_0)$ then being
of the same order of magnitude today (the so-called cosmic coincidence), the
very existence of an early universe initial big bang singularity (i.e. of an
infinite or overwhelmingly large $\dot{R}(t=0)$ in Eq. (\ref{1})) would then
oblige the initial values of these same two functions to have to obey
$\Omega_{M}(t=0) + \Omega_{\Lambda}(t=0)=1$ (no matter what the value of
$k$), to then entail that the initial conditions for Eq. (\ref{1}) would
have to be picked to incredible accuracy in order for the universe to
actually be able to evolve into one in which $\Omega_{M}(t_0)$ and
$\Omega_{\Lambda}(t_0)$ could then be of comparable magnitude today. As
such this problem is actually a variant of the original cosmological flatness
problem, with it not so much mattering that $c\Lambda$ and $\rho_M(t_0)$
would have to be of the same order of magnitude today, but rather only
that $\Lambda$ would actually be non-zero at all, since once $\Lambda$ is
non-zero, $\Omega_{M}(t_0)$ could not then be identically equal
to one today (the value it would take at all times in a $\Lambda=0$
inflationary cosmology \cite{Guth1981}), with it then being extremely
difficult to understand why $\Omega_{M}(t_0)$ had not already been redshifted
to zero this late after the big bang. Indeed, since a current era
$\Omega_{\Lambda}(t_0)=\Omega_{M}(t_0)+1/2$ universe would continue to
accelerate into the future, future observers would see a progressively
declining $\Omega_{M}(t)$ and a progressively increasing 
$\Omega_{\Lambda}(t)$, with the ensuing, no longer anywhere near close 
values of $\Omega_{M}(t)$ and $\Omega_{\Lambda}(t)$ specifically being
such as to still necessitate early universe fine tuning. While any
non-zero value for $\Lambda$ thus necessitates early universe fine tuning no
matter what that value might actually be, a universe in which
$\Omega_{\Lambda}(t_0)$ is just a little bit bigger than
$\Omega_{M}(t_0)$ today has a further peculiar feature, namely that
$\Omega_{M}(t)$ would actually be bigger than $\Omega_{\Lambda}(t)$ at all
redshifts above $z=1$ or so. Consequently, such a universe would be one
which would have decelerated continually all the time since the big bang and
which would have only started to accelerate in just our
own epoch. Since the conformal gravity theory we present below does not
possess any such switch over (its universe accelerates both below and above
$z=1$), this switch over feature of the standard theory will actually serve
both as a diagnostic for it and as way to distinguish between it and the
conformal theory, a point to which we shall return below. We thus
recognize two separate aspects to the cosmological constant problem, namely
the need to find a way to quench $\Lambda$ down from its fundamental physics
expectation in the first place, and the need to then find a way to fix
initial conditions in the early universe so as to be able to accommodate the 
non-zero quenched value for $\Lambda$ (and the thus non-unity value for
$\Omega_{M}(t_0)$) which would then ensue.

Now while a solution to the cosmological constant problem might yet be
found within standard gravity, the above described situation is so
disquieting (with the cosmological constant problem having resisted solution
for such a very long time now) as to suggest that in fact there might 
actually be something basically wrong with the whole standard picture.
Moreover, since the cosmological constant problem is a clash between two
different branches of physics, gravitational physics and particle physics,
we should not immediately assume that it is the particle physics side which
needs addressing. Rather, the indications of particle physics (actually the
better understood of the two) could well be correct, with its contribution
to $\Lambda$ actually being as big as it would appear to be, with the
problem then having to lie on the gravitational side instead. Indeed, the
very existence of the Friedmann evolution equation fine-tuning problem could
itself be an indicator that it is in fact the gravitational side which is at
fault. Thus, in the following we shall explicitly explore the implications
for cosmology of $\Lambda$ actually being a very big rather than a very
small quantity, and, with attempts to quench $\Lambda$ not having been
successful thus far, we instead turn the issue around and explore below
whether it is possible for cosmology to accommodate a large unquenched
$\Lambda$ instead. And, in particular, since the insertion of a large
$\Lambda$ into Eq. (\ref{1}) does lead to such disastrous consequences (such
as a $q(t_0)$ of order $-10^{60}$ to $-10^{120}$), we shall thus consider the
possibility that it is the standard Friedmann evolution equation which is in
need of modification. And as we shall see, its replacement below by the
cosmological evolution equation associated with the alternate conformal
gravity theory will actually lead us to a cosmology (conformal cosmology) in
which all of the above difficulties are naturally and quite readily
resolved. 

To motivate the analysis of conformal cosmology which we will present below,
we note that if we are indeed not going to quench $\Lambda$ itself, then
since it is actually $\Omega_{\Lambda}(t)$ which is its measurable
consequence for cosmology (e.g.
$q(t)=(n/2-1)\Omega_{M}(t)-\Omega_{\Lambda}(t)$), we must instead quench the
amount of gravitation which a given cosmological source produces, i.e. we
must instead quench $G$, so that instead of being controlled by the low
energy Cavendish experiment $G$, cosmology needs to instead be controlled by
an altogether smaller $G_{eff}$, with $\Omega_{\Lambda}(t)$ then being
replaced by an effective $\bar{\Omega}_{\Lambda}(t)=8\pi G_{eff} \Lambda
/3cH^2(t)$, a quantity which would then be of order one today for an
appropriately small enough $G_{eff}$. In such a case ordinary matter would
then also have a quenched coupling to cosmic evolution with
$\Omega_{M}(t)$ then being replaced by the effective
$\bar{\Omega}_{M}(t)=8\pi G_{eff} \rho_M(t)/3c^2H^2(t)$, a quantity which
would (with the standard
$\rho_{M}(t_0)$) now have to be of order $10^{-60}$, with there then being no
current era cosmic coincidence. Thus, since the
$\Omega_{\Lambda}(t)/\Omega_{M}(t)$ ratio itself is actually independent of
$G$ (so that it can generically be written as $T_V^4/T^4$ where we set
$\rho_M(t)=\sigma T^4$ for illustrative purposes), this ratio is also given
as $\bar{\Omega}_{\Lambda}(t)/ \bar{\Omega}_{M}(t)$, with the non-quenching
of $\Lambda$ then not leading us to a current era in which
$c\Lambda$ is of order $\rho_M(t_0)$. Hence in the following we will seek to
quench not the energy content of the universe but rather its effect on
cosmic evolution. Moreover, continuing in this same vein, we additionally
note that with the Friedmann equation initial condition fine-tuning problem
itself deriving from the presence of an initial big bang singularity, this
fine-tuning problem would also be avoided if there were to simply be no such
initial singularity in the first place. Thus in the following we
shall expressly look for theories in which the cosmological $G_{eff}$
is not only altogether smaller than the standard $G$, but in which it is also
even of the opposite (viz. gravitationally repulsive) sign. With such a
small, negative $G_{eff}$ both of the two aspects of the cosmological
constant problem can then readily be resolved. 
  
In order to see what would be needed of a theory which could potentially
solve the cosmological constant problem this way, it turns out to be very
instructive to first examine the problem within the context of the flat
(negligible $\Omega_{k}(t)$) inflationary universe paradigm \cite{Guth1981}.
Thus solving Eq. (\ref{1}) in the effectively flat Robertson-Walker phase
which is to follow an inflationary de Sitter era, i.e. solving for
$\rho_M(t)=B/R^3(t)$ (where $B>0$), $k=0$ and
$\Lambda>0$, yields $R(t)=(B/c\Lambda)^{1/3}sinh^{2/3}(3D^{1/2}t/2)$ where
$D=8\pi G\Lambda/3c$, so that
\begin{equation}
\Omega_{M}(t)=sech^{2}(3D^{1/2}t/2),~~\Omega_{\Lambda}(t)=D/H^2(t)
=tanh^{2}(3D^{1/2}t/2).
\label{1n}
\end{equation}
While an evolution such as this still possesses the above initial condition
fine-tuning problem for a current era $\Omega_{M}(t_0) \simeq
\Omega_{\Lambda}(t_0)$ (since the initial $\Omega_{M}(t=0) +
\Omega_{\Lambda}(t=0)$ is still equal to one), nonetheless, as regards the
magnitude of the actual contribution of the cosmological constant to cosmic
evolution, we see that no matter how large or small $\Lambda$ itself might 
be, the quantity $\Omega_{\Lambda}(t)$ itself always has to be less than or
equal to one, being given by the nicely bounded $tanh^{2}(3D^{1/2}t/2)$
form at all times. Thus given only a choice of sign for $\Lambda$, i.e. given
only that the sign is in fact the one associated with standard model
acceleration rather than deceleration, we see that the evolution
associated with Eq. (\ref{1n}) then actually keeps the magnitude of the
contribution of $\Lambda$ to cosmic evolution under control in each and
every epoch with $\Omega_{\Lambda}(t)$ always being smaller than one, and
thus never being able to be of order $10^{60}$ even if
$\Lambda$ is in fact a particle physics sized scale. (Of course in
such a case the ensuing Hubble parameter would have to readjust and be orders
of magnitude different from the currently measured one.) Cosmic acceleration
thus appears to possess the germ of a possible solution to the cosmological
constant problem within it in that it can (in principle at least) control
$\Omega_{\Lambda}(t)$. Moreover, the functional dependence
$\Omega_{\Lambda}(t)=tanh^{2}(3D^{1/2}t/2)$ given above would still be
obtained no matter what the actual numerical value of the parameter $D$,
i.e. no matter what the value of $G$. Thus if we were to replace $G$ by some
smaller $G_{eff}$ in Eq. (\ref{1}), we would still obtain a bounded form for
$\bar{\Omega}_{\Lambda}(t)= 8\pi G_{eff} \Lambda/3cH^2(t)$.

Additionally, we also note that if as well as change the magnitude
of $G$ we were also to even change its sign, so that $G_{eff}$ would then be
negative rather than positive, there would then (as already noted earlier) no
longer be any initial singularity constraint. Thus solving the standard
$k=0$ theory with the wrong sign for $G$ in Eq. (\ref{1}), but with $\Lambda$
now negative (so that $\Omega_{\Lambda}(t)$ is still positive), then yields
$R(t)=(-B/c\Lambda)^{1/3}cosh^{2/3}(3D^{1/2}t/2)$ where
$D=8\pi (-G)(-\Lambda)/3c$, so that
\begin{equation}
\Omega_{M}(t)=-cosech^{2}(3D^{1/2}t/2),~~\Omega_{\Lambda}(t)
=coth^{2}(3D^{1/2}t/2).
\label{1p}
\end{equation} 
Now, instead of being equal to unity at $t=0$, the initial
$\Omega_{M}(t=0)$ is instead infinite (since the now non-singular initial
$\dot{R}(t=0)$ vanishes rather diverges), and with the initial
$\Omega_{\Lambda}(t=0)$ being similarly infinite, initial conditions would no
longer need to be fine-tuned in order to have the associated cosmology
evolve into any particular current era value for $\Omega_{M}(t_0)$ or
$\Omega_{\Lambda}(t_0)$. This change in sign for
$G$ thus releases us from the initial $\Omega_{M}(t=0)
+\Omega_{\Lambda}(t=0)=1$ constraint and enables us to resolve the standard
model fine-tuning problem. In fact, that such a change in sign for $G$ would
immediately solve the cosmological flatness problem was already noted quite
some time ago \cite{Mannheim1992} within the context of a $\Lambda=0$
conformal cosmology, where it was pointed out that once $\Omega_{M}(t)$ is
negative (something which is the case in conformal gravity), early universe
quantities that would have had to cancel with incredible precision in Eq.
(\ref{1}) no longer need do so. With a negative cosmological $G_{eff}$ thus
readily solving the cosmological initial condition problem, it is the
purpose of the present paper to show how a small such negative $G_{eff}$ will
naturally (i.e. for a continuous, non fine-tuned range of parameters) emerge
in the alternate conformal cosmology case, to thus not only permit conformal
gravity to then readily control the contribution of
$\Lambda$ to cosmology, but to in fact do so in a way which will enable a
perfectly conventional $\rho_M(t_0)$ to contribute to cosmology according to
the non cosmic coincidence $\bar{\Omega}_{M}(t)=O(10^{-60})$ value, a value
which, despite its initial appearance, will nonetheless not in fact turn
out to be in conflict with observation. And since we have seen that there
are some potentially useful generic aspects associated with the inflationary
universe model, in order to be able to take advantage of them we turn
now to a purely kinematic, model independent study of de Sitter geometry,
one which will prove to be very instructive.

\section{Essence of the Solution to the Cosmological Constant Problem}

Guided by the discussion given above, it is very
instructive to analyze de Sitter geometry in a purely kinematic way which
requires no commitment to any particular dynamical equation of motion
\cite{Mannheim1998}, neither that of conformal gravity nor that of the
standard model either for that matter. Specifically, suppose we know only
that a given geometry is maximally 4-symmetric, i.e. that its Riemann tensor
is given by 
\begin{equation}
R^{\lambda\rho\sigma\nu}
=\alpha(g^{\sigma \rho}g^{\lambda \nu}-g^{\nu \rho}g^{\lambda\sigma}).
\label{1a}
\end{equation} 
For such a geometry contraction then yields the kinematic relation
\begin{equation} 
R^{\mu \nu} - 
g^{\mu \nu} R^{\sigma}_{\phantom {\sigma} \sigma}/2= 3\alpha g^{\mu \nu}, 
\label{1b}
\end{equation}
a relation which reduces to
\begin{equation} 
\dot{R}^2(t) +kc^2=\alpha c^2 R^2(t) 
\label{1c}
\end{equation}
when expressed in Robertson-Walker coordinates.\footnote{While Eq. (\ref{1c})
is purely kinematic, in the following it will turn out to be crucial
that the dynamical evolution equations explicitly reduce to it  
whenever the $\Omega_{M}(t)$ and $\bar{\Omega}_{M}(t)$ terms are negligible.} 
On defining $\Omega_{\Lambda}(t) =\alpha c^2 R^2(t)/\dot{R}^2(t)$
we obtain $-q(t)=\Omega_{\Lambda}(t)=1-\Omega_{k}(t)$, with $R(t)$, $q(t)$
and $\Omega_{\Lambda}(t)$ then being found \cite{Mannheim1998} to be given by

\begin{eqnarray}
R(t,\alpha<0,k<0)=(k/\alpha)^{1/2}sin((-\alpha)^{1/2}ct),
\nonumber \\
R(t,\alpha=0,k<0)=(-k)^{1/2}ct,
\nonumber \\
R(t,\alpha>0,k<0)=(-k/\alpha)^{1/2}sinh(\alpha^{1/2}ct),
\nonumber \\
R(t,\alpha>0,k=0)=R(t=0)exp(\alpha^{1/2}ct),
\nonumber \\
R(t,\alpha>0,k>0)=(k/\alpha)^{1/2}cosh(\alpha^{1/2}ct),
\label{1m}
\end{eqnarray}
and
\begin{eqnarray}  
\Omega_{\Lambda}(t,\alpha<0,k<0)=-q(t,\alpha<0,k<0)
=-tan^2((-\alpha)^{1/2}ct),
\nonumber \\ 
\Omega_{\Lambda}(t,\alpha=0,k<0)=-q(t,\alpha=0,k<0)=0, 
\nonumber \\
\Omega_{\Lambda}(t,\alpha>0,k<0)=-q(t,\alpha>0,k<0)=tanh^2(\alpha^{1/2}ct),
\nonumber \\
\Omega_{\Lambda}(t,\alpha>0,k=0)=-q(t,\alpha>0,k=0)=1,
\nonumber \\ 
\Omega_{\Lambda}(t,\alpha>0,k>0)=-q(t,\alpha>0,k>0)=coth^2(\alpha^{1/2}ct)
\label{1d}
\end{eqnarray}
in all of the various possible cases.\footnote{The $\alpha >0$ de Sitter 
geometry thus contains all three Robertson-Walker geometries and not just the
$k=0$ one, with it being the $k<0$ one which will turn out to be the most
consequential below.} As we thus see, when the parameter $\alpha$ is positive
(this actually being the case in the conformal gravity theory we study
below where $\alpha =8\pi G_{eff}\Lambda/3c^3 $ is greater than zero), each
associated solution corresponds to an accelerating universe (only the
$\alpha<0$ anti de Sitter universe decelerates), and that in each such
$\alpha>0$ universe $\Omega_{\Lambda}(t,\alpha>0)$ eventually reaches one no
matter how big the parameter $\alpha$ might be, and independent in fact of
whether or not $G$ even appears in the cosmological evolution equations at
all.\footnote{I.e. no matter how large $\alpha$ might be, the Hubble
parameter always adjusts itself to be accordingly large so that
$\Omega_{\Lambda}(t=\infty,\alpha>0)$ is then equal to one - for instance in
the familiar flat $k=0$ case $\dot{R}(t)/R(t)=\alpha^{1/2} c$ at all times.}
Moreover, while the positive spatial 3-curvature
$\Omega_{\Lambda}(t,\alpha>0,k>0)$ will only come down to one at very late
times, quite remarkably, the negative spatial 3-curvature
$\Omega_{\Lambda}(t,\alpha>0,k<0)$ will be bounded between zero and one at 
all times, no matter how large $\alpha$ might be. Thus unlike the $\alpha<0$
case where $\Omega_{\Lambda}(t)$ is unbounded, we see that when $\alpha$ is
greater than zero (the accelerating situation), $\Omega_{\Lambda}(t)$ will
either be bounded at  all times or approach a bound at late times, and that
additionally, even in the $\alpha=0$ case, the 3-curvature dominated
$\Omega_{\Lambda}(t,\alpha=0,k<0)$ will also be bounded as well. Thus we see
first, that given enough time, any accelerating de Sitter cosmology (of any
$k$) will, without any fine tuning at all, always eventually quench the
contribution of a cosmological constant to cosmology no matter how large
$\alpha$ might be, and even no matter what form the  underlying gravitational
theory might take. Second, in the particular $k<0$ case which will prove to
be the one relevant to conformal gravity itself,
$\Omega_{\Lambda}(t,\alpha>0,k<0)$ will lie below one at all times, both
early and late, and thus be bounded even at non-asymptotic times. And,
third, since $\Omega_{\Lambda}(t,\alpha=0,k<0)$ is zero, all of the $k<0$,
$\alpha \geq 0$ cosmologies will be either curvature ($R(t)\sim t$) or
cosmological constant ($R(t)\sim e^t$) dominated ones in which $0 \leq
\Omega_{\Lambda}(t) \leq 1$ at all times, to thus not only nicely constrain
the contribution of the vacuum energy density to cosmology, but to also put
it precisely into the range required by the supernovae data. We thus
distinguish between quenching $\Lambda$ and quenching $\Omega_{\Lambda}(t)$,
while noting that only the latter quenching is actually required by known
cosmological observations. Thus, the parameter $\alpha$ need not itself be
small, and in fact the larger it is, the faster
$\Omega_{\Lambda}(t,\alpha>0)$ will then approach  its asymptotic bound,
with the very same cosmic acceleration which has served to so exacerbate the
cosmological constant problem thus potentially leading to its resolution.
Having now described the essence of a possible solution to the cosmological
constant problem, we turn now to a analysis of conformal gravity itself
where this above bounding mechanism will be found to naturally appear (even
as the very presence of ordinary
$\rho_{M}(t)>0$ matter obliges the geometry to be Robertson-Walker rather
than de Sitter), with a small, negative $G_{eff}$ being found to readily
emerge and with the Cavendish experiment value for $G$ no longer so sharply
constraining cosmology.

\section{Solution to the Cosmological Constant Problem}

In attempting to depart from standard second order gravity (something we
would appear to have to do if we are indeed going to replace $G$ by
an appropriate $G_{eff}$), even within the  confines of covariant, pure
metric based theories of gravity, we immediately  realize that initially the
choice is vast, since we can in principle consider   covariant theories
based on derivative functions of the metric of arbitrarily  high order.
However, within this infinite family of higher order derivative gravitational
theories, one of them is immediately singled out, namely conformal gravity,
a fully coordinate invariant gravitational theory which possesses an
additional symmetry not enjoyed by the standard theory (viz. invariance
under any and all local conformal stretchings 
$g_{\mu\nu}(x) \rightarrow \Omega^2(x) g_{\mu\nu}(x)$ of the geometry), a
theory which consequently has as its uniquely allowed gravitational
action the Weyl action
\begin{equation} 
I_W=-\alpha_g
\int d^4x (-g)^{1/2} C_{\lambda\mu\nu\kappa}  C^{\lambda\mu\nu\kappa} 
\label{1q}
\end{equation}
where $C^{\lambda\mu\nu\kappa}$ is the conformal  Weyl tensor \cite{Weyl1918}
and where $\alpha_g$ is a purely dimensionless  gravitational coupling
constant. Conformal gravity is thus a  gravitational theory
which possess no fundamental scale (and thus no intrinsic $G$ or
fundamental $\Lambda$) at all, and is thus a theory which can immediately
lead to a cosmology which is free of intrinsic scales at sufficiently high
enough temperatures. As such, conformal gravity  emerges as a
potential gravitational analog of the Weinberg-Salam-Glashow  electroweak
theory, with it immediately being suggested
\cite{Adler1982,Zee1983} that in it Newton's constant $G$ might be
generated  as a low energy effective parameter in much the same manner as
Fermi's constant $G_F$ is generated in the electroweak theory, with
$G$ as measured in a low  energy Cavendish experiment then indeed nicely
being decoupled from the hot  early universe. However, it turns out that the
low energy limit of conformal  gravity need not emerge in precisely this
fashion since \cite{Mannheim1994} it  is not in fact necessary to
spontaneously break the conformal gravity action  down to the
Einstein-Hilbert action (i.e. down to the standard theory equations  of
motion). Rather \cite{Mannheim1994}, it is only necessary to obtain the 
solutions to those equations in the kinematic region (viz. solar system 
distance scales) where those standard solutions have been tested. Thus, as
had  been noted by Eddington \cite{Eddington1922} already in the very early
days of  relativity, the standard gravity vacuum Schwarzschild solution is
just as  equally a vacuum solution to higher derivative gravity theories 
also, since  the vanishing of the Ricci tensor entails the vanishing of its
derivatives as  well. And indeed, with variation of the conformal gravity
action leading to the  equation of motion
\cite{Dewitt1965}
\begin{equation} 
(-g)^{-1/2}\delta I_W / 
\delta g_{\mu \nu}=-2\alpha_g W^{\mu \nu}=-T^{\mu\nu}/2
\label{1e}
\end{equation}
where $W^{\mu \nu}$ is given by   
\begin{eqnarray}
 W^{\mu \nu}=g^{\mu\nu}(R^{\alpha}_{\phantom{\alpha}\alpha})   
^{;\beta} _{\phantom{;\beta};\beta}/2                                             
+ R^{\mu\nu;\beta}_{\phantom{\mu\nu;\beta};\beta}                     
 -R^{\mu\beta;\nu}_{\phantom{\mu\beta;\nu};\beta}                        
-R^{\nu \beta;\mu}_{\phantom{\nu \beta;\mu};\beta}                          
 - 2R^{\mu\beta}R^{\nu}_{\phantom{\nu}\beta}                                    
+g^{\mu\nu}R_{\alpha\beta}R^{\alpha\beta}/2 
\nonumber \\
 -2g^{\mu\nu}(R^{\alpha}_{\phantom{\alpha}\alpha})          
^{;\beta}_{\phantom{;\beta};\beta}/3                                              
+2(R^{\alpha}_{\phantom{\alpha}\alpha})^{;\mu;\nu}/3                             
+2 R^{\alpha}_{\phantom{\alpha}\alpha} R^{\mu\nu}/3                               
-g^{\mu\nu}(R^{\alpha}_{\phantom{\alpha}\alpha})^2/6,                   
\label{1f}
\end{eqnarray}
and where $T^{\mu \nu}$ is the associated energy-momentum tensor, we confirm 
immediately that the Schwarzschild solution is indeed a vacuum solution to 
conformal gravity despite the total absence of the Einstein-Hilbert action in
the purely gravitational piece $I_W$  of the conformal action. Standard 
gravity is thus seen to be only sufficient to give the standard Schwarzschild
metric  phenomenology but not at all necessary, with it thus being possible
to bypass  the Einstein-Hilbert action altogether as far as low energy
phenomena are  concerned.\footnote{Since higher derivative theories have
different  continuations to larger distances \cite{Mannheim1994}, we see that
the standard  weak gravity Schwarzschild solar system distance scale wisdom
is compatible with many  differing extrapolations to larger distances, with
standard gravity giving only  one particular possible such extrapolation. And
indeed, it has been argued \cite{Mannheim1989,Mannheim1994} that this is
actually the origin of the dark  matter problem, with standard gravity simply
giving an unsatisfactory  extrapolation to galactic distance scales and
beyond. Interestingly, the  conformal gravity extrapolation
\cite{Mannheim1997} has been found to provide  for a satisfactory explanation
of galactic rotation curve systematics without the need to introduce any
galactic dark matter at all.} As we shall show in detail  below, it is
precisely this aspect of the theory which will lead us to a  demarcation
between the high and low energy regions which is very different from that
present in the electroweak  case.\footnote{We would anyway expect there to
have to be some difference between the ways the spontaneous breakdown
mechanisms work in the two cases, since low energy gravity is not short
range.} 

While conformal gravity itself is indeed an old idea, almost as old as 
General Relativity itself in fact, it is only recently that its potential
role in cosmology and astrophysics appears to have been emphasized, with it
having been found capable \cite{Mannheim1992,Mannheim1998,Mannheim1994}
\cite{Mannheim1989,Mannheim1997,Mannheim1990,Mannheim1996,Mannheim2000}
of addressing so many of the problems (such as the flatness, horizon,
universe age, cosmic repulsion and dark matter problems) which currently
afflict standard gravity. As a fundamental theory, conformal gravity has as a
motivation the desire to give gravity a local invariance structure and a
dimensionless coupling constant (and thus power counting renormalizability),
to thereby make it analogous to the three other fundamental interactions.
And indeed, as stressed in \cite{Mannheim1990}, the local conformal symmetry
invoked  to do this then not only excludes the existence of any fundamental
mass scales  such as a fundamental cosmological constant, even after mass
scales are induced  by spontaneous breakdown of the conformal symmetry, the
(still) traceless  energy-momentum tensor then constrains any induced (and
necessarily negative) cosmological constant term  to be of the same order of
magnitude as all the other terms in $T^{\mu\nu}$,  neither smaller nor
larger. Thus, unlike standard gravity, precisely because of  its additional
symmetry, conformal gravity has a great deal of control over the 
cosmological constant (essentially, with all mass scales - of gravity and 
particle physics both - being jointly generated by spontaneous breakdown of
the scale symmetry, conformal gravity knows exactly where the zero of energy
is), and it is our purpose now to show that it is this very control which
then provides for both a natural solution to the cosmological constant
problem and for a complete accounting of the new high $z$ data.

In conformal cosmology the conformal matter action takes the form
\cite{Mannheim1992} 
\begin{equation} 
I_M=-\hbar\int d^4x(-g)^{1/2}[S^\mu S_\mu/2
- S^2R^\mu_{\phantom{\mu}\mu}/12+\lambda S^4
+i\bar{\psi}\gamma^{\mu}(x)(\partial_\mu+\Gamma_\mu(x))\psi
-gS\bar{\psi}\psi] 
\label{2}
\end{equation}
for generic massless scalar and fermionic fields, a matter action which, 
because of the selfsame underlying conformal symmetry, contains up to only
second order derivative functions of the matter fields. In this matter
action we have introduced a dimensionful scalar field $S(x)$ which is to
serve to spontaneously break the conformal symmetry,\footnote{The underlying
conformal symmetry forces the sign of the required $-S^2
R^\mu_{\phantom{\mu}\mu}/12$ term in Eq. (\ref{2}) to uniquely be negative,
and as such it would at first sight appear that after spontaneous breakdown
this term would then generate an effective low energy gravity theory which
would be repulsive. However, as we will show  below, it will turn out that
this term will in fact only be operative cosmologically where such
gravitational repulsion now appears welcome.} and for illustrative purposes
we shall first consider just one such scalar field, giving the extension
to more than one field below. And while for simplicity we use a fundamental
scalar field here, we anticipate that terms such as the $\lambda S^4$ term
will actually be generated via an effective Ginzburg-Landau theory in which
$S(x)$ is to serve as a phase transition condensate order parameter, and in
which the $\lambda S^4$ term is to represent the necessarily negative vacuum
energy density $V^{GL}_{min}= -g(T_V^2-T^2)^2$ associated with the typical
effective Ginzburg-Landau potential $V_{GL}(S,T)=gS^4- 2g(T_V^2-T^2)S^2$ at
its $T<T_V$ ordered phase minimum. In such a case the effective
$\Lambda \simeq V^{GL}_{min}$ would be negative, and would necessarily have
to in fact be so in conformal gravity, since in a theory with an underlying
conformal symmetry, the (dimensionful) vacuum energy density has to be zero
identically in the $S(x)=0$ scale invariant high temperature regime above
all scale generating phase transitions where the presence of an exact,
unbroken conformal symmetry ensures the absence of any fundamental
scale. Thus unlike the situation in the standard theory, in the conformal
theory the sign of $\Lambda$ is actually known a priori, in fact explicitly
being opposite to that assumed in the standard model. However, despite being
negative (something we simulate below by taking the parameter $\lambda$ in
Eq. (\ref{2}) to be negative), as we shall see, it will precisely be this
particular sign which will actually lead to cosmic acceleration in the
conformal theory.

For the above matter action the matter field equations of motion take the
form  
\begin{eqnarray}
 i\gamma^{\mu}(x)[\partial_{\mu} +\Gamma_\mu(x)]\psi 
- g S \psi =0,  
\nonumber \\
S^\mu _{\phantom{\mu};\mu}
+ SR^\mu_{\phantom{\mu}\mu}/6-
4\lambda S^3+ g\bar{\psi}\psi=0,
\label{2a}
\end{eqnarray}
with the matter energy-momentum tensor being given by 
\begin{eqnarray}
T^{\mu \nu}=\hbar \{i \bar{\psi} \gamma^{\mu}(x)[ 
\partial^{\nu}                    
+\Gamma^\nu(x)]\psi
+2S^\mu S^\nu/3 
-g^{\mu\nu}S^\alpha S_\alpha/6 
- SS^{\mu;\nu}/3
\nonumber \\             
+ g^{\mu\nu}SS^\alpha_{\phantom{\alpha};\alpha}/3                              
- S^2(R^{\mu\nu}-
g^{\mu\nu}R^\alpha_{\phantom{\alpha}\alpha}/2)/6
-g^{\mu \nu}\lambda S^4 \}.  
\label{2b}
\end{eqnarray}
Thus in the case where the scalar field acquires a non-zero vacuum
expectation value (an expectation value which can always be rotated into a
spacetime constant $S_0$ by an appropriate local conformal transformation),
the entire energy-momentum tensor of the theory is found (for a perfect
matter fluid $T^{\mu\nu}_{kin}$ of the fermions) to take the form         
\begin{equation}
T^{\mu\nu}=T^{\mu\nu}_{kin}-\hbar S_0^2(R^{\mu\nu}-
g^{\mu\nu}R^\alpha_{\phantom{\alpha}\alpha}/2)/6            
-g^{\mu\nu}\hbar\lambda S_0^4;
\label{3}
\end{equation}
with the complete solution to the scalar, fermionic and gravitational field
equations of motion in a background Robertson-Walker geometry (viz. a
geometry in which the Weyl tensor and $W^{\mu \nu}$ both  vanish) then
reducing \cite{Mannheim1998} to just one relevant equation, namely 
\begin{equation}
T^{\mu\nu}=0,
\label{3a}
\end{equation}
a remarkably simple condition which immediately fixes the zero of energy, so
that, and unlike the situation in the standard theory, the conformal theory
does indeed know exactly where the zero of energy is. Moreover, not only
does this condition fix the zero of energy, it even serves to fix the spatial
curvature of the universe as well, with Eq. (\ref{3a}) itself reducing to
the even simpler $T_{kin}^{\mu\nu}=0$ at the earliest temperatures above all
phase transitions, a condition which is then only satisfiable non-trivially
\cite{Mannheim2000} when $k<0$, with the positive energy of ordinary matter
being explicitly canceled by the negative gravitational energy associated
with negative spatial curvature. Thus in the conformal theory the global
topology of the universe is fixed once and for all before any
cosmological phase transition ever occurs. 

With the imposition of Eq. (\ref{3a}) enabling us to rewrite
Eq. (\ref{3}) in the form  
\begin{equation}
\hbar S_0^2(R^{\mu\nu}-
g^{\mu\nu}R^\alpha_{\phantom{\alpha}\alpha}/2)/6 =T^{\mu\nu}_{kin}           
-g^{\mu\nu}\hbar\lambda S_0^4,
\label{3aa}
\end{equation}
we thus see that the evolution equation of conformal cosmology looks 
identical to that of standard gravity save only that the quantity 
$-\hbar S_0^2 /12$ has replaced the familiar $c^3/16 \pi G$. With this 
homogeneous and isotropic, global scalar field $S_0$ filling all space and 
acting cosmologically, we see that this change in sign compared with
standard  gravity leads to a cosmology in which gravity is globally
repulsive rather  than attractive. Because of this change in sign, 
conformal cosmology thus has no initial singularity (i.e. it expands from  a
finite minimum radius), and is thus precisely released from the standard
big  bang model fine-tuning constraints described earlier. Similarly, because
of this change in  sign the contribution of $\rho_{M}(t)$ to the expansion of
the universe is  now effectively repulsive, to (heuristically) mesh with the
phenomenological high $z$ data fits in which $\Omega_{M}(t)$ was allowed to
go negative. Apart from a  change in sign, we see that through $S_0$ there
is also a change in the  strength of gravity compared to the standard
theory. It is this feature which  will prove central to the solution to the
cosmological constant  problem which we present below.

Despite the fact that conformal gravity has now been found to be globally 
repulsive, nonetheless, it is important to note that in the conformal theory 
local solar system gravity can still be attractive; with it having been 
specifically found \cite{Mannheim1994} that for a static, spherically 
symmetric source such as a star, the conformal gravity field equation of Eq.
(\ref{1e})  reduces to a fourth order (i.e. not a second order) Poisson
equation $\nabla ^4 g_{00}=3(T^0_{\phantom{0} 0} -T^r_{\phantom{r}
r})/4\alpha_g  g_{00}\equiv -f(r)$, with solution 
$-g_{00}(r)= 1-2\beta^*/r+\gamma^* r$ where $\beta^{*}=\int dr f(r) r^4/12$ 
and $\gamma^{*}=-\int dr f(r) r^2/2$. With the coupling constant $\alpha_g$
in the  Weyl action $I_W$ simply making no contribution in highly symmetric 
cosmologically relevant geometries where $C^{\lambda\mu \nu \kappa}$ and  
$W^{\mu \nu}$ vanish, and with the sign of $\beta^*$  being directly given by 
the sign of this thus cosmologically irrelevant $\alpha_g$, we see that 
locally attractive and globally repulsive gravity are now decoupled and thus
able to coexist. Local gravity is thus fixed by local sources alone, sources
which are only gravitational inhomogeneities in the otherwise homogeneous
global cosmological background, i.e. sources which are characterized by
small, local variations in the background scalar field $S(x)$, variations
which themselves are completely decoupled from the homogeneous, constant,
cosmological background field $S_0$ itself. It is thus the distinction
between homogeneity and inhomogeneity which provides the demarcation between
local and global gravity, to thus now enable us to consider repulsive
cosmologies which are not incompatible with the attractive gravity observed
on solar system distance scales.

Given the equation of motion $T^{\mu \nu}=0$, the conformal cosmology 
evolution equations are then found to take the form (on setting 
$\Lambda=\hbar\lambda S^4_0$)   
\begin{equation}
\dot{R}^2(t) +kc^2 =
-3c^3\dot{R}^2(t)(\Omega_{M}(t)+
\Omega_{\Lambda}(t))/ 4 \pi \hbar S_0^2 G 
\equiv \dot{R}^2(t)(\bar{\Omega}_{M}(t)+
\bar{\Omega}_{\Lambda}(t))
\label{4}
\end{equation}
\begin{equation}
\bar{\Omega}_{M}(t)+ \bar{\Omega}_{\Lambda}(t)+\Omega_{k}(t)=1,~~
q(t)=(n/2-1)\bar{\Omega}_{M}(t)-\bar{\Omega}_{\Lambda}(t)
\label{1j}
\end{equation}
where Eq. (\ref{4}) serves to define $\bar{\Omega}_{M}(t)$ and
$\bar{\Omega}_{\Lambda}(t)$. As we see, precisely because the underlying
conformal invariance has forced the conformal $T^{\mu \nu}$ to be of the
standard second order form, Eq. (\ref{4}) is found to be remarkably similar
in form to Eq. (\ref{1}), with conformal cosmology thus only containing
familiar ingredients. As an alternate  cosmology then, conformal gravity
thus gets about as close to standard gravity  as it is possible for an
alternative to get while nonetheless still being different. In fact
the two sets of evolution equations are actually completely equivalent to
each other save only in the replacement of the standard attractive $G$ by a
repulsive effective cosmological $G_{eff}$ given by $-3c^3/4\pi \hbar
S_0^2$, a replacement which precisely gives the conformal $G_{eff}$ the very
structure desired above of an effective cosmological $G_{eff}$ which is
to indeed solve the cosmological constant problem. With such a $G_{eff}$ we
also see that the larger $S_0$ the smaller the resulting contribution of
$\Lambda$ to cosmic evolution, with conformal gravity thus being able to
accommodate a far larger $\Lambda$ than the standard theory. 

In order to see whether conformal gravity can actually accommodate the new
supernovae data with such a $G_{eff}$, it is necessary to analyze
the solutions to Eq. (\ref{4}). Such solutions are readily obtained
\cite{Mannheim1998}, and can be classified  according to the signs of
$\lambda$ and $k$ (and even though we have indicated above that
$\lambda$ and $k$ are to both be negative in the conformal case,
nonetheless for completeness we explore all possible cases here). In the
simpler to treat high temperature era where $\rho_{M}(t)=A/R^4=\sigma T^4$
the complete family of solutions is given as 
\begin{eqnarray}
R^2(t,\alpha<0,k<0)=k(1-\beta)/2\alpha+
k\beta sin^2 ((-\alpha)^{1/2} ct)/\alpha,
\nonumber \\
R^2(t,\alpha=0,k<0)=-2A/k\hbar c S_0^2-kc^2t^2,
\nonumber \\
R^2(t,\alpha>0,k<0)= -k(\beta-1)/2\alpha
-k\beta sinh^2 (\alpha^{1/2} ct)/\alpha,
\nonumber \\
R^2(t,\alpha > 0,k=0)=(-A/\hbar\lambda c S_0^4)^{1/2}
cosh(2\alpha^{1/2}ct),
\nonumber \\
R^2(t,\alpha > 0,k>0)=k(1+\beta)/2\alpha+
k\beta sinh^2 (\alpha^{1/2} ct)/\alpha,
\label{5}
\end{eqnarray}
where we have introduced the parameters $\alpha =-2\lambda S_0^2$ and 
$\beta =(1- 16A\lambda/k^2\hbar c)^{1/2}$. Similarly the associated 
deceleration parameters take the form
\begin{eqnarray}
q(t,\alpha<0,k<0)=
tan^2((-\alpha)^{1/2}ct)
-2(1-\beta)cos(2(-\alpha)^{1/2}ct)/
\beta sin^2(2(-\alpha)^{1/2}ct),
\nonumber \\
q(t,\alpha=0,k<0)=-2A/k^2\hbar c^3 S_0^2t^2,
\nonumber \\
q(t,\alpha>0,k<0)=
-tanh^2(\alpha^{1/2}ct)
+2(1-\beta)cosh(2\alpha^{1/2}ct)/
\beta sinh^2(2\alpha^{1/2}ct),
\nonumber \\
q(t,\alpha>0,k=0)=-1-2/sinh^2(2\alpha^{1/2}ct),
\nonumber \\
q(t,\alpha>0,k>0)=
-coth^2(\alpha^{1/2}ct)
-2(1-\beta)cosh(2\alpha^{1/2}ct)/
\beta sinh^2(2\alpha^{1/2}ct).
\label{6}
\end{eqnarray}
Now while Eq. (\ref{5}) yields a variety of temporal behaviors for $R(t)$, it 
is of great interest to note that every single one of them begins 
with $\dot{R}(t=0)$ being zero rather than infinite (with 
$\bar{\Omega}_{M}(t=0)$ and $\bar{\Omega}_{\Lambda}(t=0)$ then both being
infinite regardless of what their current era values might be), and that
each one of the solutions in which $\lambda$ is negative (viz. $\alpha>0$)
is associated with a universe which permanently expands (only the $\lambda
>0$ solution can recollapse, with conformal cosmology thus  correlating the
long time behavior of $R(t)$ with the sign of $\lambda$  rather than with
the sign of
$k$). We thus need to determine the degree to  which the permanently
expanding universes have by now already become  permanently accelerating.

To this end we note first from Eq. (\ref{6}) that with $\beta$ being greater 
than one when $\lambda$ is negative, both the $\alpha >0,~k<0$ and the 
$\alpha >0,~k=0$ cosmologies are in fact permanently accelerating ones no 
matter what the values of their parameters. To explore the degree to which 
they have by now already become asymptotic, as well as to
determine the acceleration properties of the $\alpha >0,~k>0$ cosmology, we 
note that since each of the solutions given in Eq. (\ref{5})
has a non-zero minimum radius, each associated $\alpha >0$ cosmology has some 
very large but finite maximum temperature $T_{max}$ given by
\begin{eqnarray}
T_{max}^2(\alpha>0,k<0)/T^2(t,\alpha>0,k<0)= 1
+2\beta sinh^2 (\alpha^{1/2} ct)/(\beta-1),
\nonumber \\
T_{max}^2(\alpha > 0,k=0)/T^2(t,\alpha>0,k=0)=
cosh(2\alpha^{1/2}ct),
\nonumber \\
T_{max}^2(\alpha > 0,k>0)/T^2(t,\alpha>0,k>0)=
1+2\beta sinh^2 (\alpha^{1/2} ct)/(\beta+1),
\label{7}
\end{eqnarray}
with all the permanently expanding ones thus necessarily being way below 
their maximum temperatures once given enough time. To obtain further 
insight into these solutions it is convenient to introduce an effective 
temperature according to $-c\hbar \lambda S^4_0=\sigma T_V^4$. In terms of 
this $T_V$ we then find that in all the $\lambda<0$ cosmologies the energy 
density terms take the form
\begin{eqnarray}
\bar{\Omega}_{\Lambda}(t)=(1-T^2/T_{max}^2)^{-1}(1+T^2T_{max}^2/T_V^4)^{-1},~
\nonumber \\
\bar{\Omega}_M(t)=-(T^4/T_V^4)\bar{\Omega}_{\Lambda}(t),~
\label{8}
\end{eqnarray}
where $(\beta-1)/(\beta+1)=T_V^4/T_{max}^4$ for the $k<0$ case, and where 
$(\beta-1)/(\beta+1)=T_{max}^4/T_V^4$ for the $k>0$ case. With $\beta$ being
greater than one, we find that for the $k>0$ case $T_V$ is greater than 
$T_{max}$, for $k=0$ $T_V$ is equal to $T_{max}$, and for $k<0$ $T_V$ is less 
than $T_{max}$, with the energy in curvature (viz. the energy in the 
gravitational field itself) thus making a direct contribution to the maximum 
temperature of the universe. Hence, simply because the temperature $T_{max}$ 
is overwhelmingly larger than the current temperature $T(t_0)$ (i.e. simply 
because the universe has been expanding and cooling for such a long time now), 
we see that, without any fine tuning at all, in both the $k>0$ and $k=0$ cases 
(i.e. cases where $T_V \geq T_{max} \gg T(t_0)$), the quantity 
$\bar{\Omega}_{\Lambda}(t_0)$ is already at its asymptotic limit of one today, 
that $\bar{\Omega}_M(t_0)$ is completely suppressed, and that the deceleration 
parameter is given by $q(t_0)=-1$. 

For the $\alpha>0$, $k<0$ case (the only $\alpha>0$ case where $T_V$ is less 
than $T_{max}$, with a large $T_V$ thus automatically entailing an even
larger $T_{max}$) a very different outcome is possible however. Specifically,
since in this case the quantity $(1+T^2T_{max}^2/T_V^4)^{-1}$ is always 
bounded between zero and one no matter what the relative magnitudes of
$T_V$, $T_{max}$  and $T(t)$, we see that as long as $T_{max} \gg T(t_0)$
(which would even have to be the case if $T_V \gg T(t_0)$), rather than
having had to have already reached its asymptotic limit of one by now, the
quantity $\bar{\Omega}_{\Lambda}(t_0)$ is instead only required to be 
bounded by it. And with it thus expressly being bounded from above, in the
$k<0$ case the current era value of $\bar{\Omega}_{\Lambda}(t_0)$ thus has
to lie somewhere between zero and one today no matter how big or small $T_V$
might be (and despite the fact that $\bar{\Omega}_{\Lambda}(t_0)$ is even
infinite in the early universe); with the simple additional
requirement that $T_V$ also be very much greater than $T(t_0)$ (viz. large
$\Lambda$) then entailing that $\bar{\Omega}_M(t_0)$ will yet again be
completely suppressed in the current era. Moreover, 
$\bar{\Omega}_{\Lambda}(t_0)$ will take a typical value of one half
should the value of the quantity $T^2(t_0)T_{max}^2/T_V^4$  currently be
close to one. Values of $\bar{\Omega}_{\Lambda}(t_0)$ not merely less than
one but even appreciably so are thus readily achievable in the $k<0$ case
for a continuous range of temperature parameters which obey  $T_{max}\gg T_V
\gg T(t_0)$ without the need for any fine-tuning at all. And with the
current era $\bar{\Omega}_M(t_0)$ being completely suppressed and with
the early universe $\bar{\Omega}_{M}(t=0)$ being infinite, there is
thus no cosmic coincidence fine-tuning problem either, with conformal
cosmology being completely free of fine-tuning problems. 

Noting from Eq. (\ref{7}) that the temporal evolution of the $\alpha >0$, 
$k<0$ case is given by
\begin{equation}
tanh^2 (\alpha^{1/2} ct)=(1-T^2/T_{max}^2)/(1+T^2T_{max}^2/T_V^4),
\label{7a}
\end{equation}
we see that simply because of the fact that $T_{max} \gg T(t_0)$, the current
era value of $\bar{\Omega}_{\Lambda}(t_0)$ is then given by the nicely
bounded $tanh^2(\alpha^{1/2}ct_0)$ form, i.e. given precisely by the
form found in the  model independent analysis of de Sitter space that was
presented  above. Additionally, in this case $\Omega_k(t_0)$ is then given by
$sech^2(\alpha^{1/2}ct_0)$, with negative spatial curvature then explicitly
contributing to current era cosmology (and even doing so repulsively 
since $q(t_0)=(n/2-1)(1+kc^2/\dot{R}^2(t_0))
-n\bar{\Omega}_{\Lambda}(t_0)/2)$), with the current era universe not needing
to yet be vacuum (viz. cosmological constant) dominated.
Robertson-Walker conformal cosmology thus goes through three epochs,
matter domination (where $q(t) \sim -\infty$), curvature domination
($R(t)\sim t$, $q(t)\sim 0$) and then finally vacuum domination ($R(t)\sim
e^t$, $q(t) \sim -1$), with matter domination of the expansion rate being
restricted to the very early universe (thus incidentally making it
irrelevant for later epochs whether we take $\rho_{M}(t)=A/R^4$ or
$\rho_{M}(t)=B/R^3$ in them), with the current era universe lying somewhere
in or between the curvature and vacuum dominated phases. The universe can
thus currently be in a mild linearly expanding phase, with it not needing to
have to already be in a far more rapidly growing exponential one.
Additionally, we also note that in the
$\alpha>0$,
$k<0$ case the Hubble parameter is given by
\begin{equation}
H(t)=\alpha^{1/2}c(1-T^2(t)/T^2_{max})/tanh(\alpha^{1/2}ct), 
\label{7b}
\end{equation}
with its current value thus obeying $-q(t_0)H^2(t_0)=\alpha c^2$ and with the
current age of the universe being given by $t_0 H(t_0)= arctanh
[(-q(t_0))^{1/2}]/(-q(t_0))^{1/2}$.\footnote{The age $t_0$ is thus
necessarily greater than $1/H(t_0)$ ($t_0=1/H(t_0)$  when $\alpha=0$), with
it taking the typical value $t_0=1.25/H(t_0)$ when $q(t_0)=-1/2$. Thus as
already noted in \cite{Mannheim1996,Mannheim1998} conformal cosmologies have
no universe age problem.} Thus no matter what the explicit value of
$\alpha$ the current era $q(t_0)$ always has to lie between zero and minus
one with the Hubble parameter adjusting itself so as to be given by
$H^2(t_0)=-\alpha c^2/q(t_0)$, and with the $-q(t)H^2(t)$ product being
independent of time at all times $T(t) \ll T_{max}$. The quantity
$\bar{\Omega}_{\Lambda}(t)H^2(t)$ thus asymptotes to $\alpha c^2$, a
behavior quite reminiscent of that found in the inflationary universe
which we exhibited earlier as Eq. (\ref{1n}). Thus when $k$ is negative
(as noted earlier this is actually the theoretically preferred choice in the
conformal theory), we find that values of $\bar{\Omega}_{\Lambda}(t_0)$
less than one are then naturally achievable in the conformal theory, and with
$\bar{\Omega}_{\Lambda}(t_0)$ not being able to be larger than one (no
matter what the value of $k$ in fact) once given only that the parameter
$\alpha$ is in fact positive and that the universe is as old as it is, we
see that a conformal cosmology universe solves the cosmological constant
problem simply by living for a very long time.  

Thus we see that in all three of the $\alpha >0$ cases the simple requirement 
that $T_{max} \gg T(t_0)$, $T_{V} \gg T(t_0)$ ensures that 
$\bar{\Omega}_M(t_0)$ is completely negligible at current temperatures (it 
can thus only be relevant in the early universe), with the current era Eq. 
(\ref{4}) then reducing to 
\begin{equation}
\dot{R}^2(t) +kc^2 =
\dot{R}^2(t)\bar{\Omega}_{\Lambda}(t)
\label{10}
\end{equation}
to thus not only yield as a current era conformal cosmology what in the
standard theory could only possibly occur as a very late  one, but to also
yield one which enjoys all the nice purely kinematic  properties of a de
Sitter geometry which we identified above. Since studies of  galaxy counts
indicate that the purely visible matter contribution to $\Omega_{M}(t_0)$ is
of order one (actually of order $10^{-2}$ or so in  theories in which dark
matter is not considered), it follows from Eq. (\ref{4}) that current era
suppression of $\bar{\Omega}_{M}(t_0)$ will in fact be achieved if the
conformal cosmology scale parameter $S_0$ is altogether larger than the
inverse Planck length $L^{-1}_{PL}$, a condition which is naturally 
imposable (i.e. for a continuous, non fine-tuned, range of parameters of the 
theory) and which is precisely compatible with a large rather than a small
$S_0$. Comparison with Eq.  (\ref{1}) shows that a current era $\lambda<0$
conformal cosmology looks exactly like a low mass standard model cosmology,
except that instead of $\Omega_{M}(t_0)$ being negligibly small (something
difficult to understand in the standard theory) it is
$\bar{\Omega}_{M}(t_0)=-3\Omega_{M}(t_0)/4\pi S_0^2 L^2_{PL}$ which is
negligibly small instead ($\Omega_{M}(t_0)$ itself need not actually be
negligible in conformal gravity - rather, it is only the contribution of
$\rho_{M}(t)$ to the evolution of the current universe which needs be
small).  Hence, we see that the very essence of our work is that the same
mechanism which causes $\bar{\Omega}_{\Lambda}(t_0)$ to be of order one
today, viz. a large rather than a small $\Lambda=\hbar\lambda S^4_0$, serves
at the same time, and without any fine tuning, to cause
$\bar{\Omega}_{M}(t_0)$ to decouple from current era cosmology, and to thus
not have to take a value anywhere near close in magnitude to that taken by
$\bar{\Omega}_{\Lambda}(t_0)$. Thus to conclude we see that when
$\lambda$ is negative, that fact alone is sufficient to automatically drive
us into the narrow (supernovae data compatible) $\bar{\Omega}_{M}(t_0)=0$,
$0\leq \bar{\Omega}_{\Lambda}(t_0)\leq 1$ window,\footnote{Given the
similarity of Eqs. (\ref{1}) and (\ref{4}), phenomenological high
$z$ data fits are thus fits to both standard and conformal gravity, with
conformal gravity nicely locking into an $\bar{\Omega}_{M}(t_0)=0$ window in
which a perfectly normal $\rho_{M}(t)$ makes a completely negligible
contribution to cosmic expansion.} with the current era
$\bar{\Omega}_{\Lambda}(t_0)$ being able to be less than (and even much less
than) one in the negative spatial curvature case.

While we have thus shown that in conformal gravity it is indeed possible to
completely control the contribution of the cosmological constant to cosmic
evolution without the need for any fine-tuning, for practical applications
of the theory it would be helpful if we could constrain the allowed values
of the parameter $\alpha$ which controls the actual bounded
$tanh(\alpha^{1/2}ct_0)$ value which $\bar{\Omega}_{\Lambda}(t_0)$ takes.
Additionally, we also need to extend our analysis to allow for the presence
of more than one scalar field, and it is thus of interest to note that it is
just such an extension which will actually enable us to provide the desired
$\alpha$ constraint. Specifically, since conformal invariance is to be
completely unbroken in the very earliest universe with the vacuum energy
density $\Lambda=V_{min}^{GL}$ being zero at temperatures above the (highest)
critical temperature $T_V$ (a temperature which we recall is less than
$T_{max}$ in the $k<0$ case), and since the $R^2(t,\alpha=0,k<0)=-2A/k\hbar
c S_0^2-kc^2t^2$ scale factor has a non-zero minimum value even in the
absence of any $\alpha$ (with Eq. (\ref{4}) actually only admitting of
negative $k$ solutions when $\Lambda$ is zero), we thus see that in the
conformal theory there would be a maximum (negative curvature supported)
temperature $T_{max}^2\sim -k\hbar c S_0^2/2A$ even in the absence of
spontaneous breakdown of the conformal symmetry, with the values of
$T_{max}$ and $T_V$ thus being completely decoupled.\footnote{In passing we
note that in a $\Lambda=0$ conformal cosmology $q(t)$ is found
\cite{Mannheim1996} to be given by $q(t)=[1-T_{max}^2/T^2(t)]^{-1}$), with
its current era value thus being negligibly small. Thus, as had actually
been noted well in advance of the recent high $z$ supernovae data,
$\Lambda=0$ conformal cosmology already possesses a repulsion not present in
a $\Lambda=0$ standard model (where $q(k=0,t_0)=1/2$).} In order to actually
have a non-zero $S_0$ above all critical temperatures, i.e. to actually have
a non-zero $S_0$ even without any spontaneous breakdown,  we must thus
introduce a fundamental, non-dynamical, conformally coupled "urfeld" scalar
field $S(x)$ which fills all space. For such a field, the solution to its
field equation $S^{\mu}_{\phantom{\mu}\mu}+ SR^{\mu}_{\phantom{\mu}\mu}/6=0$
in a maximally 3 symmetric background geometry would have to depend only on
time, with a purely time dependent conformal transformation then bringing it
to a constant value $S_0$, with a resetting of the time bringing the
simultaneously conformally transformed geometry back to the standard
Robertson-Walker form, and with $R^2(t,\alpha=0,k<0)$ then being given by
Eq. (\ref{5}) just as required. In addition to this urfeld we introduce at a
much lower temperature a second scalar field, one associated with a typical
particle physics symmetry breaking vacuum expectation value $S_0^{\prime}$
where $S_0^{\prime}$ is thus altogether smaller than $S_0$. While both of
these fields will conformally couple to the Ricci scalar in Eq. (\ref{2}),
it will be the much larger early universe urfeld $S_0$ which will dominate
$G_{eff}$ (so that our requirement that  $S_0$ be altogether larger than
$L^{-1}_{PL}$ is thus readily satisfiable even while $S_0^{\prime}$ itself is
typically much smaller), while it will be the particle physics one which
will generate $-c\Lambda=-c\hbar \lambda^{\prime} S_0^{\prime 4}=\sigma
T_V^4$. In the  presence of both of these fields we can continue to use the
previous formalism provided we identify the effective $\lambda$ parameter in
Eqs. (\ref{2}) and (\ref{3}) according to $\lambda=\lambda^{\prime}
S_0^{\prime 4}/S_0^4 \ll\lambda^{\prime}$. With  such a choice $T_{max}/T_V$
will typically be as large as $S_0/S_0^{\prime}$ with the parameter $\alpha
c^2=-2\lambda S_0^2=-2\lambda^{\prime}S_0^{\prime 4}/S_0^2$ then being much
smaller than $-2\lambda^{\prime}S_0^{\prime 2}$. Thus with $q_0$ being given
by $-tanh^2(\alpha^{1/2}ct_0) = -(1+T^2(t_0)T_{max}^2/T_V^4)^{-1}$, the
quantity $\alpha^{1/2}ct_0$ can thus readily be small enough to prevent $q_0$
from already being at its asymptotic value of minus one, with a really big
$T_{max}$ bringing $q(t_0)$ very close to its curvature dominated value
of zero.

Now while we have already indicated that $k<0$ spatial curvature is
theoretically preferred in conformal gravity, we also note that there is
even observational  support for this value \cite{Mannheim1997}, support
obtained from an at first highly unlikely source,  namely galactic rotation
curve data. Recalling that in conformal gravity  the metric outside of a
static spherically symmetric source such as a  star is given by
$-g_{00}(r)= 1-2\beta^*/r+\gamma^* r$, we see that in the  conformal theory
the departure from Newton is found to be given by a potential  that actually
grows (linearly) with distance. Hence, unlike the situation in  standard
gravity, in conformal gravity it is not  possible to ever neglect the matter
exterior to any region of interest, with  the rest of the universe (viz. the
Hubble flow) then also contributing to  galactic motions (i.e. a test
particle in a galaxy not only samples the local  galactic gravitational
field, it also samples that of the global Hubble  flow as well). And indeed,
it was found
\cite{Mannheim1997} that the effect  on galaxies of the global Hubble flow
was to generate an additional linear  potential with a universal coefficient
given by
$\gamma_0/2=(-k)^{1/2}$, i.e. one which is generated explicitly by the
negative scalar curvature of the universe\footnote{Essentially, under the
general coordinate transformation $r=\rho/(1-\gamma_0 \rho/4)^2$, $t = \int
d\tau / R(\tau)$, a static,  Schwarzschild coordinate observer in the rest
frame of a given galaxy  recognizes the (conformally transformed) comoving
Robertson-Walker metric $ds^2=\Omega(\tau, \rho)[c^2 d\tau^2 - R^2(\tau)
(d\rho^2 + \rho^2 d\Omega)/ (1-\rho^2\gamma_0^2/16)^2]$ (where $\Omega(\tau,
\rho)=  (1+\rho\gamma_0/4)^2/ R^2(\tau)(1-\rho\gamma_0/4)^2$) as being
conformally  equivalent to the metric 
$ds^2=(1+\gamma_0 r)c^2dt^2-dr^2/(1+\gamma_0 r)-r^2d\Omega$.} 
(heuristically, the repulsion associated with negative scalar curvature 
pushes galactic matter deeper into any given galaxy, an effect which an
observer inside that galaxy interprets as attraction), with conformal gravity
then being found able to give an acceptable accounting of galactic rotation 
curve systematics (in the data fitting $\gamma_0$ is numerically found to be 
given by $3.06 \times 10^{-30}$ cm$^{-1}$, i.e. to be explicitly given by a 
cosmologically significant length scale) without recourse to dark matter at 
all. We thus identify an explicit imprint of cosmology on galactic rotation 
curves, recognize that it is its neglect which may have led to the need for 
dark matter, and for our purposes here confirm that $k$ is indeed 
negative.\footnote{Given the presence of the imprint of such a cosmological 
scale on galaxies, it thus becomes necessary (see also 
\cite{McGaugh1998,Mannheim2000} for  related discussion) for dark matter
models to  equally produce such a scale, something which may not be all that
easy in  standard flat $k=0$ models where no curvature scale is available.}
Conformal  cosmology thus leads us directly to $\bar{\Omega}_{M}(t_0)=0$, 
$\bar{\Omega}_{\Lambda}(t_0)=tanh^2(\alpha^{1/2}ct_0)$, and would thus appear 
to lead us naturally right into the region favored by the new high $z$ data.

As regards the role played by negative $\Lambda$ in conformal cosmology, we 
recall that the effect of elementary particle physics phase transitions is 
to lead to a vacuum energy density $V_{min}^{GL}$ which is typically
expected  to be negative rather than positive since each one of the many
particle  physics phase transitions acts to lower the vacuum energy density
some more as the  universe cools. Once given such a negative $\Lambda$ its
effect on cosmic evolution then depends on the sign of the effective $G$.
Thus, for repulsive conformal cosmology, negative $\Lambda$  translates into
positive $\bar{\Omega}_{\Lambda}(t)$ and thus to cosmic  acceleration,
whereas for the attractive standard cosmology it translates into  negative
$\Omega_{\Lambda}(t)$, to then not lead to any cosmic acceleration at  all.
Thus added to the challenges faced by the standard theory is the need to
explain not only why $\Lambda$ should be small, but also to explain why it 
should also not in fact be negative, with the very fact of cosmic
acceleration providing some support for the central theme of our work,
namely that  cosmologically, the effective 
$G$ is in fact negative.

As regards such an effective negative cosmological $G$, we note that are 
essentially two primary arguments which have in the past supported the 
contrary, $G$ positive, cosmological position, namely the current value of 
$\Omega_{M}(t_0)$ and big bang nucleosynthesis. However, of these two, the
$\Omega_{M}(t_0)$ argument now has to be discounted. Specifically, with 
earlier (i.e. pre high $z$) data having led to a current value of 
$\Omega_{M}(t_0)=8\pi G\rho_{M}(t_0)/3c^2H^2(t_0)$ which was tantalizingly 
close to one (provided one included dark matter that is), it strongly 
suggested that cosmology was indeed normalized to the gravitational constant
$G$, with  cosmological theory otherwise having to explain this 
closeness as an accident. And, indeed, the great appeal of inflation was that 
it provided a rationale for having $\Omega_{M}(t_0)$ be close to one today by 
having $\Omega_{M}(t)$ be identically equal to one in each and every epoch.
However, with the new high $z$ data, we now know that $\Omega_{M}(t_0)$ is
unambiguously less than one, and more, that it will get ever smaller as the
universe continues to accelerate. Thus, for observers  sufficiently far
enough into the future $\Omega_{M}(t)$ will be nowhere near  one, with its
current closeness to one being only an artifact of the particular epoch in
which current observers happen to be making observations (and of  course
without dark matter, by itself known explicitly detected luminous  matter
only yields for $\Omega_{M}(t_0)$ a value which is actually a few orders  of
magnitude or so below one today).

With regard to nucleosynthesis, we note that, in principle, it only requires 
that the universe had once been hot enough to have been able to trigger 
nuclear reactions, with it not being at all necessary that even earlier there
had been  an altogether hotter ($G>0$ induced) big bang phase. And indeed, it
has been found \cite{Knox1993,Elizondo1994,Lohiya1999} that since the
universe has  been expanding and cooling for such a very long time now,
conformal cosmology  is also capable of having once been hot enough to have
undergone  nucleosynthesis (even without a big bang); with the latest
calculations \cite{Lohiya1999} yielding  the requisite amount of helium as
well as the metallicity which is explicitly  seen in population II
stars,\footnote{Even though the cosmology expands far  more slowly than the
standard one, nonetheless, this gets compensated  for in the conformal
case  because weak interactions are then  found \cite{Lohiya1999}
to remain in thermal equilibrium down to lower temperatures than in the
standard case, with the conformal metallicity  predictions then being found
\cite{Lohiya1999} to actually outperform those of  the standard model.} with
the inability \cite{Knox1993,Elizondo1994,Lohiya1999}  of conformal cosmology
to yield a sufficient amount of deuterium being its only  outstanding
nucleosynthesis problem. Now, as regards the production of  deuterium, we
note that while it is generally thought difficult to produce
post-primordially, this is not quite the case, as it is actually fairly easy
to  both produce and then retain deuterium by spallation or fragmentation of
light  nuclei \cite{Hoyle1973,Epstein1976,Epstein1977}, particularly if the
spallation  is pre rather than post galactic.\footnote{Even though
spallation models of  deuterium production were never actually ruled out,
the models were quickly set  aside once it became apparent that deuterium
could be  produced by standard big bang nucleosynthesis.} In fact the
problem is then not  one of an underproduction of deuterium, but rather of
an overproduction of the  other light elements. However, as noted by Epstein
\cite{Epstein1977}, if the  spallation is to also take place in the early
universe with its onset occurring after the nucleosynthesis itself, then (i)
in such a situation only hydrogen  and helium interactions would be of any
significance, with only $Z \leq 3$  nuclei then being producible, and (ii)
that in such a case the  high energies involved would serve to favor
deuterium production over the  lithium production which is favored at
ordinary energies. In addition to this, the authors of \cite{Lohiya1999}, 
on having found the helium abundance in their nucleosynthesis calculations
to be a rather sensitive function of the baryon to entropy ratio, have
suggested that lithium production could also be  suppressed if the spallation
were to take place inhomogeneously with  helium deficient clouds then
spallating with helium rich ones. In such a case,  deuterium would then be
produced not during nucleosynthesis itself but some  time afterwards just as
inhomogeneities first begin to form in the  universe.\footnote{It could thus
be of interest to measure the lithium to  deuterium abundance ratio of high
$z$ quasar absorbers, with the obtaining of a  value for this ratio different
from that expected in standard big bang  nucleosynthesis then possibly
indicating the occurrence of inhomogeneous but  still fairly early universe
spallation.} Since a theory for the growth of inhomogeneities in conformal
cosmology has not yet been developed, unfortunately, it is not yet possible 
to currently provide a detailed analysis of this issue or assess its
implications for conformal gravity, though the issue is of course of
paramount importance for the conformal theory.   

While nucleosynthesis continues to be the primary achievement of standard
big bang cosmology, nonetheless, quite recently, with the advent of precision
measurements of the anisotropy of the cosmic microwave background on very
small angular scales \cite{deBernardis2000}, the standard theory has 
actually run into a potential problem, with it proving to be somewhat
difficult \cite{Tegmark2000} to fit the anisotropy data (and particularly
their apparent lack of any second acoustic peak) using the baryon density
which is explicitly inferred from big bang nucleosynthesis.\footnote{While 
the standard theory had always been able to fit the light element abundances
with a very specific phenomenological value for the baryon density (as
determined solely from the abundance fitting itself), we note in passing
that there had never actually been any independent confirmation of that
particular value.} And while it is of course far too early to draw any
permanent conclusions, nonetheless the current situation could be another
indicator, albeit so far only a mild one, that $G$ might not in fact control
cosmology.

Also, of course, by the very same token, these selfsame anisotropy data also
provide a new challenge to alternate gravitational theories as well, with it
therefore being somewhat urgent to determine the specific anisotropy
predictions of the conformal theory. And with conformal gravity being
confined to the rather tight $\bar{\Omega}_{M}(t_0)=0$,  $0 \leq
\bar{\Omega}_{\Lambda}(t_0) \leq 1$,
$1 \geq \Omega_{k}(t_0) \geq 0$ window (a window which obliges
$\bar{\Omega}_{M}(t)$ to still be negligible at recombination), such
predictions may well prove to be definitive for the theory.\footnote{These
numbers should not be  compared with the currently preferred standard model
$\Omega_{M}(t_0)\approx 0.3$, $\Omega_{\Lambda}(t_0) \approx 0.7$,
$\Omega_{k}(t_0) \approx 0$  numbers (see e.g. \cite{deBernardis2000}),
since the extraction of those numbers from data requires the assumption of a
dynamical early universe fluctuation model (as well as the use, of course,
of a completely mysterious and not understood value for
$\Omega_{\Lambda}(t_0)$). It is for that reason that we have focused on  the
supernovae data in the present paper, since we can extract information from
them (such as a sign and magnitude for $q(t_0)$) without recourse to any
particular dynamical model at all.} Thus again it becomes necessary to
develop a theory for the growth of conformal cosmology inhomogeneities.
(Essentially, with the global cosmic background being homogeneous, and with
local gravity only arising through inhomogeneities in the conformal gravity
theory, galaxy fluctuation theory sits just at the point where local and
global conformal effects become competitive.)

While conformal gravity can thus be definitively tested at recombination,
it is important to note that there is actually a much nearer by redshift at
which definitive testing can also be made, namely \cite{Mannheim2001} that
provided by only a modest extension of the Hubble plot a little beyond
$z=1$. Specifically, we had noted earlier that the standard model actually
predicts that the universe be decelerating above $z=1$ (a thus definitive
test for it in and of itself) while the conformal gravity theory continues
to be accelerating above $z=1$ ($q(t,\alpha>0,k<0)$ of Eq. (\ref{6}) is
always negative); with Hubble plot data at around $z=2$ (where the conformal
theory is dimmer than the standard theory by about $0.3$ magnitudes) thus
enabling one to make a fairly definitive discrimination between the two
theories. Specifically, in \cite{Mannheim2001} it was shown that the
luminosity distance redshift relation associated with $\alpha>0$, $k<0$
conformal gravity is given as
\begin{equation}
H(t_0) d_L/c=-(1+z)^2\left\{1-[1+q(t_0)-q(t_0)/(1+z)^2]^{1/2}\right\}/q(t_0),
\label{11}
\end{equation}
with the ensuing conformal gravity fits to the $z<1$ data presented in
\cite{Mannheim2001} being found to be every bit as good as the corresponding 
$\Omega_{M}(t_0)= 0.3$, $\Omega_{\Lambda}(t_0) = 0.7$ standard model fits.
Specifically, for 54 fitted $z<1$ data points of \cite{Perlmutter1998}
and \cite{Hamuy1996}, $q(t_0)=-0.37$ conformal gravity gives $\chi^2=58.6$,
$q(t_0)=0$ conformal gravity gives $\chi^2=61.5$, while
$\Omega_{M}(t_0)= 0.3$, $\Omega_{\Lambda}(t_0) = 0.7$ standard gravity gives
$\chi^2=57.7$, with the predictions of the two theories thus being
completely indistinguishable below $z<1$, and with one therefore needing to
go above $z=1$ to discriminate. Thus despite the longstanding conviction that
ordinary matter is a substantial contributor to cosmology in the nearby
universe, available $z<1$ data are just as compatible with ordinary matter
making no contribution at all (viz. $\bar{\Omega}_{M}(t_0)=0$). 

Now it would be quite remarkable if two totally different theories could
account for the very same $z<1$ data, and it is thus worthwhile to identify
why this occurs. To this end we note that all $\Omega_{k}(t_0)=0$ standard
model fits have to lie between its ($\Omega_{M}(t_0)= 1$,
$\Omega_{\Lambda}(t_0)=0$) and ($\Omega_{M}(t_0)= 0$,
$\Omega_{\Lambda}(t_0)=1$) limits, limits for which the respective
luminosity functions are given by the (too bright for the data)
$d_L=cH(t_0)^{-1}2(1+z)[1-(1+z)^{-1/2}]$ and the (too dim)
$d_L=cH(t_0)^{-1}(z+z^2)$, and for which the respective $\chi^2$ for the 54
$z<1$ data points are found to be $\chi^2=92.9$ and $\chi^2=75.8$. Now while
both of these $\chi^2$ values are sufficiently far from the data so as to
exclude them, they are not overwhelmingly far from the data (since $z$ does
not get large enough to cause the various expectations for $d_L$ to differ
that much), thus making it possible for some intermediate (not too bright,
not too dim) prediction such as ($\Omega_{M}(t_0)= 0.3$,
$\Omega_{\Lambda}(t_0)=0.7$) to then work. Similarly, the
$\Omega_{k}(t_0)=1$ (viz. $q(t_0)=0$) conformal gravity prediction (the
not too bright, not too dim $d_L=cH(t_0)^{-1}(z+z^2/2)$) also does not differ
from them that much either at $z<1$ allowing it to work well too. The data
below $z=1$ can thus be accounted for by the universe being near to either
$\Omega_{k}(t_0)=0$ or to $\Omega_{k}(t_0)=1$, values for $\Omega_{k}(t_0)$
which are not orders of magnitude apart, with it being necessary to
go to higher $z$ to discriminate. Thus it could be that the standard model is
correct and that the $\Omega_{k}(t_0)=1$ conformal gravity fit is
fortuitous, or it could be that conformal gravity is correct with it then
being the $\Omega_{k}(t_0)=0$ standard model fit which is the fortuitous one,
with this issue then readily being resolved as soon as the Hubble plot can be
definitively pushed beyond $z=1$.  

Also conformal gravity can of course equally be tested at altogether higher
redshifts such as at those associated with the cosmic microwave background
and nucleosynthesis, and while one should not understate the seriousness of
the inhomogeneity and deuterium problems for conformal cosmology (indeed its
very viability as a cosmological model is contingent upon a successful
resolution of these very issues), nonetheless, the relative ease with which
conformal gravity deals with cosmological constant problem, the most severe
problem the standard theory faces, would appear to entitle the conformal
theory to further consideration. And even if the conformal gravity 
alternative were to fall by the wayside, nonetheless our analysis of the
role  that $G$ plays in engendering the standard model cosmological constant
problem would still  remain valid. Moreover, the cosmological constant
problem is not the only problem which besets standard gravity, with dark
matter theory itself having steadily accumulated a whole host of unresolved
challenges of its own \cite{Sellwood2000}, especially in the area of
galactic dynamics. And indeed, after identifying more than 10 distinct such
problems, the authors of \cite{Sellwood2000} venture to suggest that it is
no longer obvious that the formidable difficulties facing alternate theories
are any more daunting than those facing dark matter.      

To conclude this paper, we note once again that spontaneous breakdown effects 
such as those associated with a Goldstone boson pion or with massive 
intermediate vector bosons seem to be very much in evidence in current era 
particle physics experiments, and are thus not quenched at all apparently. 
Hence all the evidence of particle physics is that its contribution to 
$\Lambda$ should in fact be large rather than small today. However, since in 
such a case it is nonetheless possible for $\bar{\Omega}_{\Lambda}(t_0)$ to 
still be small today, we see that the standard gravity fine tuning problem 
associated with having $\Omega_{M}(t_0) \simeq \Omega_{\Lambda}(t_0)$ today 
can be viewed as being not so much one of trying to understand why it is 
$\Omega_{\Lambda}(t_0)$ which is of order one after 15 or so billion years, 
but rather of trying to explain why the matter density contribution to
cosmology  should be of order one after that much time rather than a factor
$T^4/T_V^4$  smaller. Since this latter problem is readily resolved if $G$
does not in fact  control cosmology, but if cosmology is instead controlled
by some altogether smaller length squared scale such as $G_{eff}=-3c^3/4\pi
\hbar S_0^2$, we see that the origin of the entire cosmological constant
problem can be directly traced to the assumption that gravity is controlled
by Newton's constant $G$ on each and every distance scale; with the very
existence of the cosmological constant problem possibly being an indicator
that the extrapolation of standard gravity from its solar system origins
all the way to cosmology might be a lot less reliable than is commonly
believed. 

The author wishes to thank Drs. D. Lohiya and M. Sher for useful
discussions. This work has  been supported in part by the Department of
Energy under grant No.  DE-FG02-92ER40716.00.

\end{document}